\documentclass[preprint,12pt]{elsarticle}
\usepackage{amssymb}
\usepackage{amsmath}
\usepackage{multirow}
\usepackage{array}

\journal{Physica B: Condensed Matter}

\begin{document}

\begin{frontmatter}

\author[msu]{V. Torlao}
\ead{torlao.vc752@s.msumain.edu.ph}

\author[msu]{E. A. Fajardo\corref{cor1}}
\ead{edwardaris.fajardo@msumain.edu.ph}

\affiliation[msu]{organization={Department of Physics},
            addressline={Mindanao State University - Main Campus}, 
            city={Marawi City},
            postcode={9700}, 
            state={Lanao del Sur},
            country={Philippines}}

\cortext[cor1]{Corresponding author}

\title{Formation Energy Prediction of Material Crystal Structures using Deep Learning}

\begin{abstract}
Determining the stability of chemical compounds is essential for advancing material discovery. In this study, we introduce a novel deep neural network model designed to predict a crystal's formation energy, which identifies its stability property. Our model leverages elemental fractions derived from material composition and incorporates the symmetry classification as an additional input feature. The materials' symmetry classifications represent the crystal polymorphs and are crucial for understanding phase transitions in materials. Our findings demonstrate that the integration of crystal system, point group, or space group symmetry information significantly enhances the predictive performance of the developed deep learning architecture, where the highest accuracy was achieved when space group classification was incorporated. In addition, we use the same model architecture to predict the energy above hull, an indicator to material stability, with formation energy as an additional input feature.
\end{abstract}


\end{frontmatter}

\section{Introduction}
\label{sec:intro}

Formation energy is a fundamental property that quantifies the energy change during the formation of a compound from its constituent elements, indicating whether energy is absorbed or released in the process \cite{kirklin2015open}. This property exhibits an underlying connection to material stability. A compound is typically considered thermodynamically stable if its formation energy lies on or below the convex hull, as this suggests that the material is the most energetically favorable configuration among possible phases. Conversely, if the formation energy is above the convex hull, the compound is typically regarded as unstable with the exception of metastable materials whose formation energies are slightly above the hull and can be stable or synthesizable under specific conditions \cite{sun2016thermodynamic}.

The widespread availability of materials databases such as Materials Project \cite{jain2013commentary} and Open Quantum Materials Database \cite{saal2013materials, kirklin2015open} have opened an opportunity for machine learning to be used for accelerated materials discovery and in prediction of material properties such as formation energy. This includes models for formation energy calculations that inputs material attributes, chemical compositions, or crystallographic classifications using machine learning algorithms based on Bayesian statistics \cite{ward2018matminer}, aritificial \cite{ye2018pauling, jha2018elemnet} and convolutional \cite{goodall2020roost, jain2018atomic} neural networks, and decision trees \cite{ward2017including, kim2018machine}.

Here, we adopt the featurized chemical formula approach of Jha et al. \cite{jha2018elemnet}, called the ElemNet model, which uses chemical elemental fractions as input feature. One key challenge to this model is the presence of materials with the same chemical formula but exist in different structural phases, also known as the crystal polymorphs. Our aim is to tackle this multi-class problem by adding symmetry labels to the various chemical systems. These descriptors serve as our input features in a deep learning model that predicts formation energy. Furthermore, we use the same model architecture, but with the addition of each material's formation energy as an input feature, to predict the energy above the hull and identify if a material is stable. Ultimately, our goal is to predict the formation energy of chemical compounds across various space groups, allowing us to identify the specific space group in which a compound can potentially form its most stable structure.

This paper is organized as follows. Section \ref{sec:data-mining} covers data mining, which includes details on the dataset, structure of the input features, and specifications of the deep learning model. The results and discussion section is found in Sec.\ \ref{sec:results-discussions} which shows comparisons of the formation energy prediction models with and without symmetry classifications in the input features. Additionally, this section presents the results of energy above hull predictions for Manganese-Nickel-Oxygen chemical systems. Finally, Sec.\ \ref{sec:conclusions} shows the conclusions.

\section{Data Mining}
\label{sec:data-mining}

Using the Materials Project database \cite{jain2013commentary}, we extracted a comprehensive dataset comprising 153,232 material entries. Of these, 104,351 are theoretically derived data, calculated using Density Functional Theory (DFT), while 48,884 are experimentally observed. 

It is noteworthy that some materials share the same chemical composition but are categorized under different crystallographic classifications, including crystal systems, point groups, and space groups. Among the total number of material entries, only 104,037 distinct chemical compositions are listed. Additionally, there are 122,303 distinct entries, 128,068 distinct entries, and 133,260 distinct entries for materials with crystal system, point group, and space group classifications, respectively. This is summarized in Table~\ref{tab:crystallographic_classifications}.

\begin{table}[h]
  \centering
  \setlength{\tabcolsep}{10pt}
  \renewcommand{\arraystretch}{1.5}
  \begin{tabular}{lc}
    \hline \hline
    classification type & distinct entries \\
    \hline
    distinct chemical compositions & 104,037 \\
    crystal system classification & 122,303 \\
    point group classification & 128,068 \\
    space group classification & 133,260 \\
    total number of entries & 153,232 \\
    \hline \hline
  \end{tabular}
  \caption{Breakdown of distinct chemical compositions and their crystallographic classifications. This table shows the number of distinct chemical compositions and their corresponding crystallographic classifications, including crystal systems, point groups, and space groups, within the dataset.}
  \label{tab:crystallographic_classifications}
\end{table}

\subsection{Data preprocessing}

The chemical compounds in the extracted data consist of combinations taken from 86 chemical elements, as the remaining elements from the periodic table do not form stable materials. This process results in a dataframe with 86 columns, each representing an element. The extracted data also includes symmetry information, with a column for the designated symmetry classification under a specific crystallographic category. The symmetry classifications are then converted into binary format using one-hot encoding, resulting in 7 columns for crystal system classification, 32 columns for point group classification, and 228 columns for space group classification (while there are 230 space groups, two do not exist in the Materials Project data.) The final dataframe contains both elemental columns and one-hot encoded columns for crystal system, point group, or space group classfications as input features, with formation energy as the target feature. The data is then split into an 80 percent training set and a 20 percent testing set.

\subsection{Deep Learning}

\begin{figure}[htbp]
	\centering
	\includegraphics[scale=0.26]{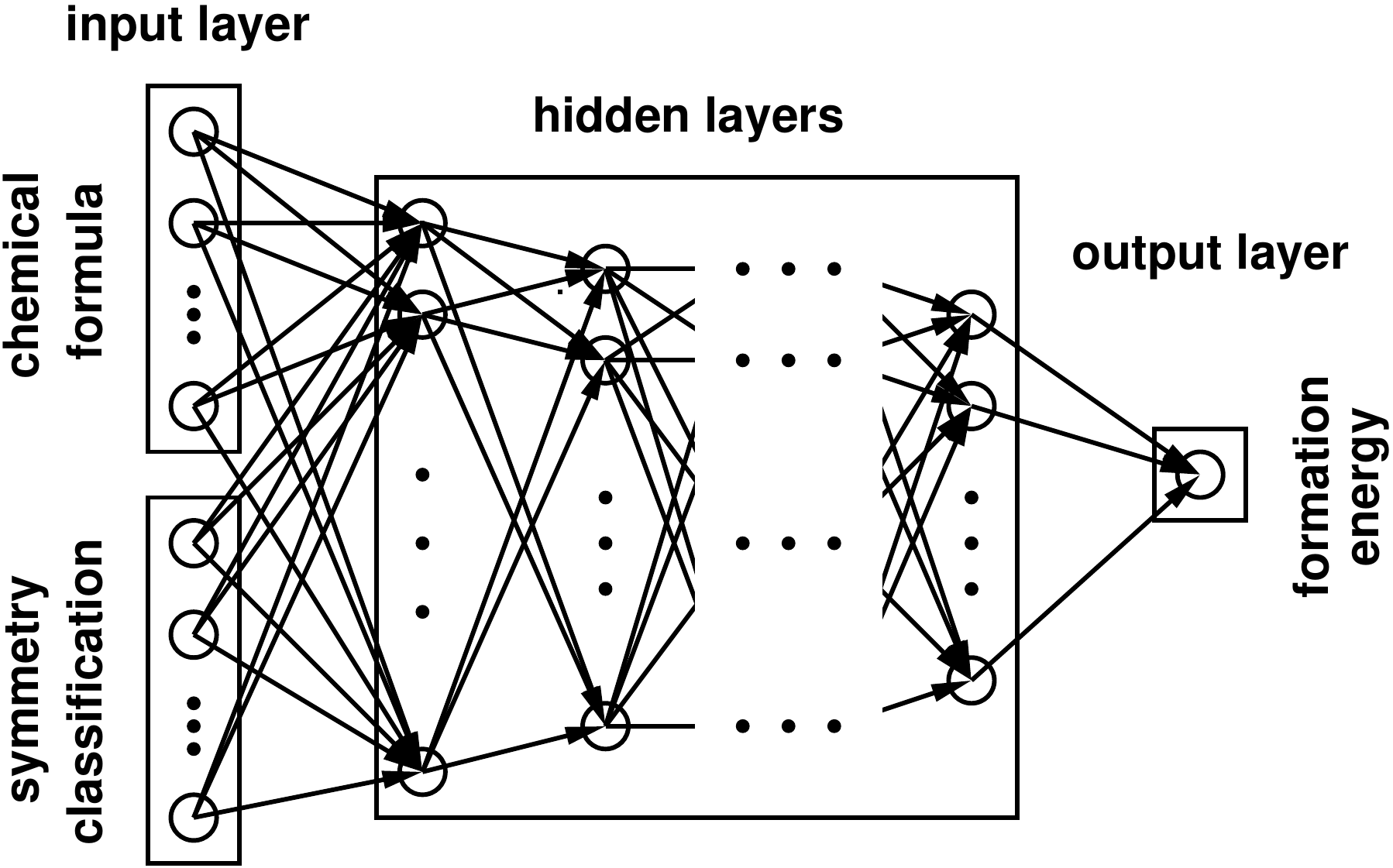}
	\caption[Neural Network Diagram]{Visualization of the utilized neural network architecture composed of an input layer, hidden layers, and a single neuron output layer all connected by weights. The input layer consists of information about the chemical formula and a symmetry classification such as crystal system, point group, or space group.}
	\label{fig:nn}
\end{figure}

Complementary to theoretical and experimental approaches, machine learning algorithms such as deep learning can be used to directly predict formation energy. Formation energy is a crucial indicator of material stability, and deep learning offers a robust method for making such predictions. Deep learning leverages neural networks \cite{aggarwal2018neural}, which are sophisticated artificial systems inspired by the structure and function of biological neural networks. A visualization of an artificial neural network is shown in FIG.\ \ref{fig:nn}. Within these networks, information traverses through multiple hidden layers. Each layer processes its inputs using weights and biases before generating an output. This computational process can be represented mathematically as
\begin{equation}
\hat{y} = f\left(\sum_{j=1}^{d} w_j x_j + b\right),
\label{eq:neural_network}
\end{equation}
where $\hat{y}$ is the output, $f$ is the activation function, $w_j$ are the weights, $x_j$ are the $d$ total number of inputs, and $b$ is the bias term. In the network, each neuron in the hidden layers computes the value based on Eq.\ \ref{eq:neural_network}, processing its received input values from the previous layer. This computed output is then fed into neurons in the subsequent layer, continuing this flow until the final output neuron is reached, which produces the ultimate output value. This output is compared against the target value, and the network's parameters are adjusted via an optimizer to minimize the error. This iterative process continues until the error is reduced to a satisfactory level.

For the hidden layers, we utilized the Rectified Linear Unit (ReLU)\cite{glorot2011deep} activation function, defined as
\begin{equation}
\text{ReLU}(x) = \max(0, x).
\end{equation}
This function outputs $x$ if it is positive; otherwise, it outputs zero. The output layer employs a linear activation function.

To optimize the network, we employed the Adam (Adaptive Moment Estimation) \cite{kingma2015adam} optimizer, which updates the network's weights and biases by combining the benefits of AdaGrad (Adaptive Gradient Algorithm) \cite{duchi2011adaptive} and RMSProp (Root Mean Square Propagation) \cite{tieleman2012root}. From AdaGrad, Adam takes the concept of adjusting the learning rate for each parameter based on the historical gradients. However, instead of accumulating all past gradients like AdaGrad, Adam uses RMSProp's idea of an exponentially decaying moving average of squared gradients to avoid the learning rate decaying too quickly.

\subsection{Deep Learning Architecture}
\label{subsec:deep-learning-architecture}

The deep neural network model comprises six sequential hidden layers, with the first and second hidden layers containing 512 neurons each, followed by a third layer with 256 neurons, a fourth layer with 128 neurons, a fifth layer with 64 neurons, and a sixth layer with 32 neurons. ReLU activation functions were applied to all hidden layers. The output layer utilizes a Linear activation function, suitable for regression tasks where the network directly predicts continuous values such as the prediction of formation energies. The optimizer being used is Adam. The maximum number of epochs is set at 500. However, to avoid overfitting, we implement early stopping with a patience parameter of 10 epochs, which is the number of epochs to tolerate no improvement. The dataset was used to train the deep neural network model. Following training, the model was tested and evaluated using Mean Absolute Error (MAE), Mean Squared Error (MSE) and Root Mean Squared Error (RMSE) as the loss function metrics. Additionally, the R-squared $(R^2)$ value was computed to assess the model's predictive confidence.

\section{Results and Discussions}
\label{sec:results-discussions}

\subsection{Chemical Formula as Input Feature}

\begin{figure}[htbp]
	\centering
	\includegraphics[scale=0.4]{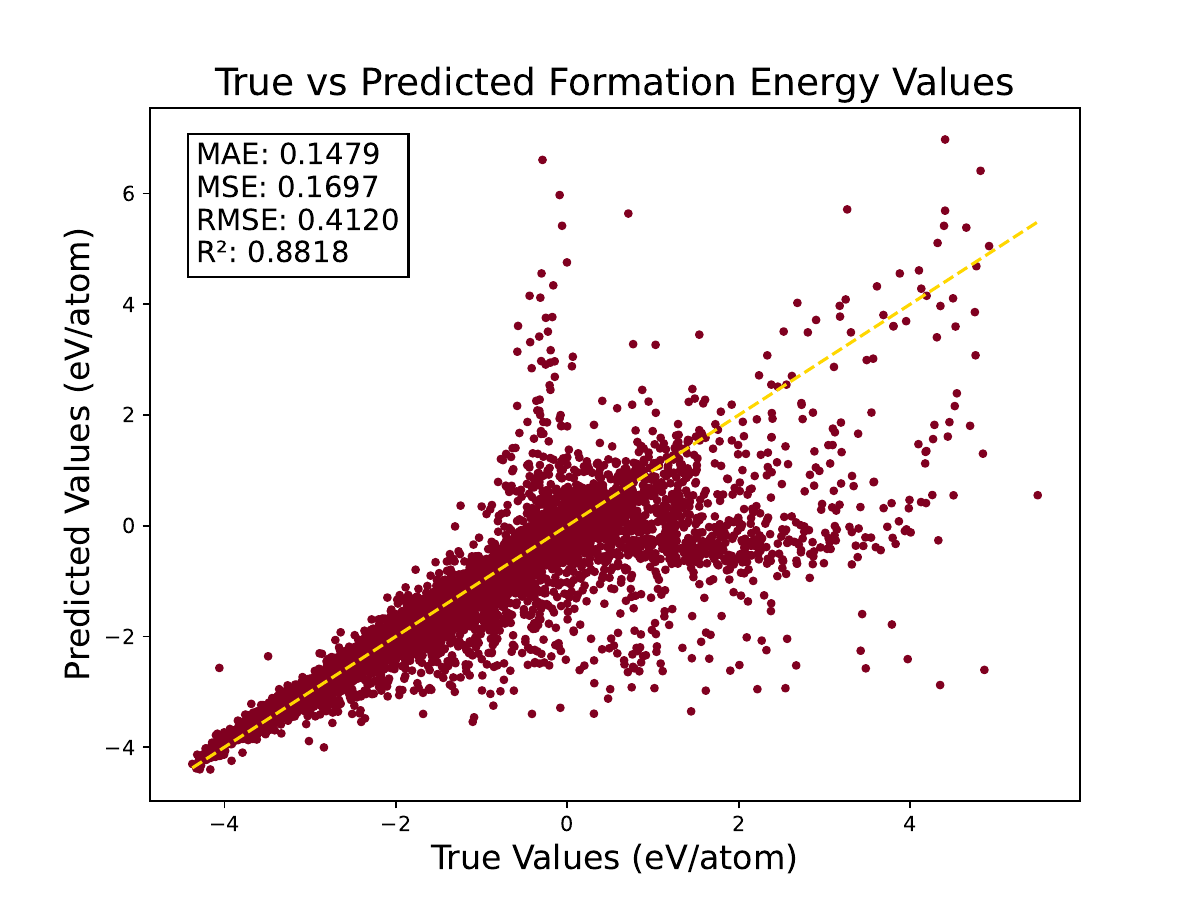}
	\caption{True values vs.\ predicted formation energies using chemical formula only as input. This figure shows the plot of true values vs.\ predicted values of formation energy using all entries in the dataset with chemical formula only as input. The calculated error metrics are as follows: MAE = 0.1479 eV/atom, MSE = 0.1697 eV/atom, RMSE = 0.4120 eV/atom and $R^2$ = 0.8818.}
	\label{fig:cc}
      \end{figure}

In this research, we derived our primary approach from ElemNet \cite{jha2018elemnet}, which utilizes elemental fractions from chemical formulas, sourced from the OQMD, as input features to predict formation enthalpy. Inspired by this, we trained a deep learning model using the architecture described in Sec.\ \ref{subsec:deep-learning-architecture}, but with Materials Project data. In ElemNet, the model's performance is enhanced by excluding single elements and outliers pertaining to the formation enthalpy values outside $\pm5$ standard deviations. In our approach we retained the single elements and removed data outside $\pm7$ standard deviations from the formation energy of the Materials Project data. This adjustment led to the removal of three materials, leaving a total of 153,229 training and testing data points. The training process terminated at epoch 34 due to the early stopping condition. The plot of true values vs.\ predicted values are shown in FIG.\ \ref{fig:cc}, and reflects a reasonable error.

      We now train the same deep learning architecture using distinct chemical formula only. For duplicated chemical formulas, i.e.\ materials with the same chemical formulas but have different formation energy values due possibly to their varying structures, we keep only for each chemical formula that one with the lowest formation energy. The number of dataset used in training and testing is now the same as the total number of distinct formula, which is 104,037 as mentioned in Table~\ref{tab:crystallographic_classifications}. The training for this case terminated after 38 epochs. The plot for true values vs.\ predicted values shown in FIG.\ \ref{fig:2x2}(a) depicts a better result than the previously trained model that contains duplicated entries of chemical formula. This is evident from the lower error metrics MAE, MSE, and RMSE and higher R$^2$.

\subsection{Inclusion of Symmetry Characterization}

Given that materials data often contains a material exhibiting different symmetries, the same material can appear multiple times across various classifications such as crystal systems, point groups, and space groups. A single material may exhibit different classifications under varying conditions or due to subtle structural differences, such as variations in unit cell volume or density. By utilizing these symmetry classifications, we can enhance the feature set of the materials and reduce instances where a single material with a specific symmetry label is assigned multiple formation energy values. In such instances, we utilize only the material entry with the smallest formation energy.

\begin{figure}[htbp]
	\centering
	\includegraphics[scale=0.45]{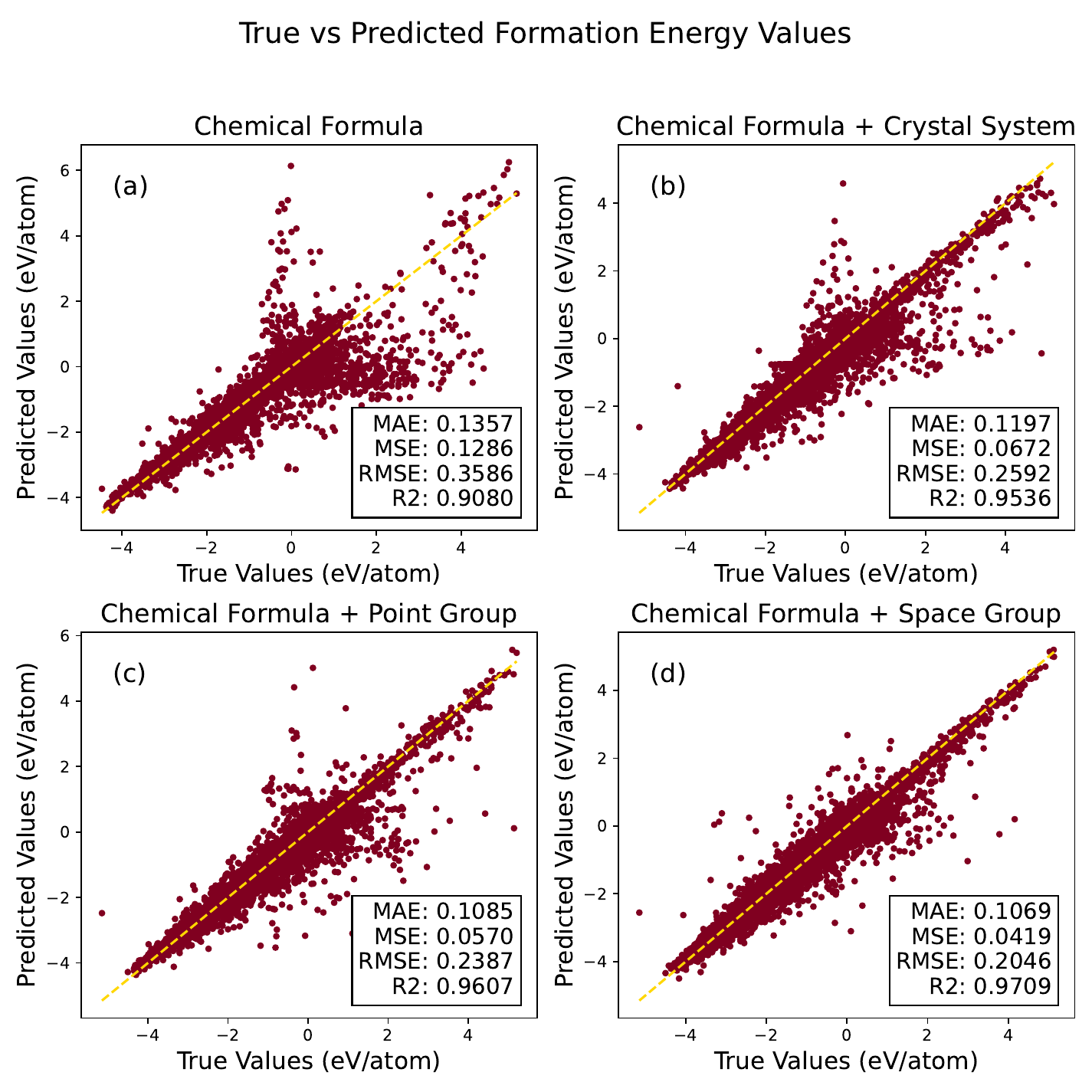}
	\caption{True values vs.\ predicted formation energies using chemical formula and symmetry classifications as input. This figure shows the plots of true values vs.\ predicted values of formation energy and corresponding error metrics for dataset with input features containing materials of distinct (a) chemical formula only, (b) crystal system classifications, (c) point group classifications, and (d) space group classifications. In all cases, only the lowest formation energy is utilized for each duplicated input feature.}
	\label{fig:2x2}
\end{figure}

By removing duplicated input features for symmetry characterized materials, the amount of dataset used for training and testing is the same as the number of distinct entries for each symmetry classification type displayed in Table~\ref{tab:crystallographic_classifications}. With early stopping implemented, the total number of epochs in the training process is 51, 73, and 51, corresponding respectively to the models with materials characterized under crystal system, point group, and space group symmetry classifications. As shown in FIGs.\ \ref{fig:2x2}(b)-\ref{fig:2x2}(d), incorporating these symmetry classifications to the input feature generated improved predictive accuracies. The inclusion of space group information yields the highest accuracy improvement, followed by point group information, and then crystal system information. 

\subsection{Energy above hull prediction}

\begin{figure}[htbp]
	\centering
	\includegraphics[scale=0.4]{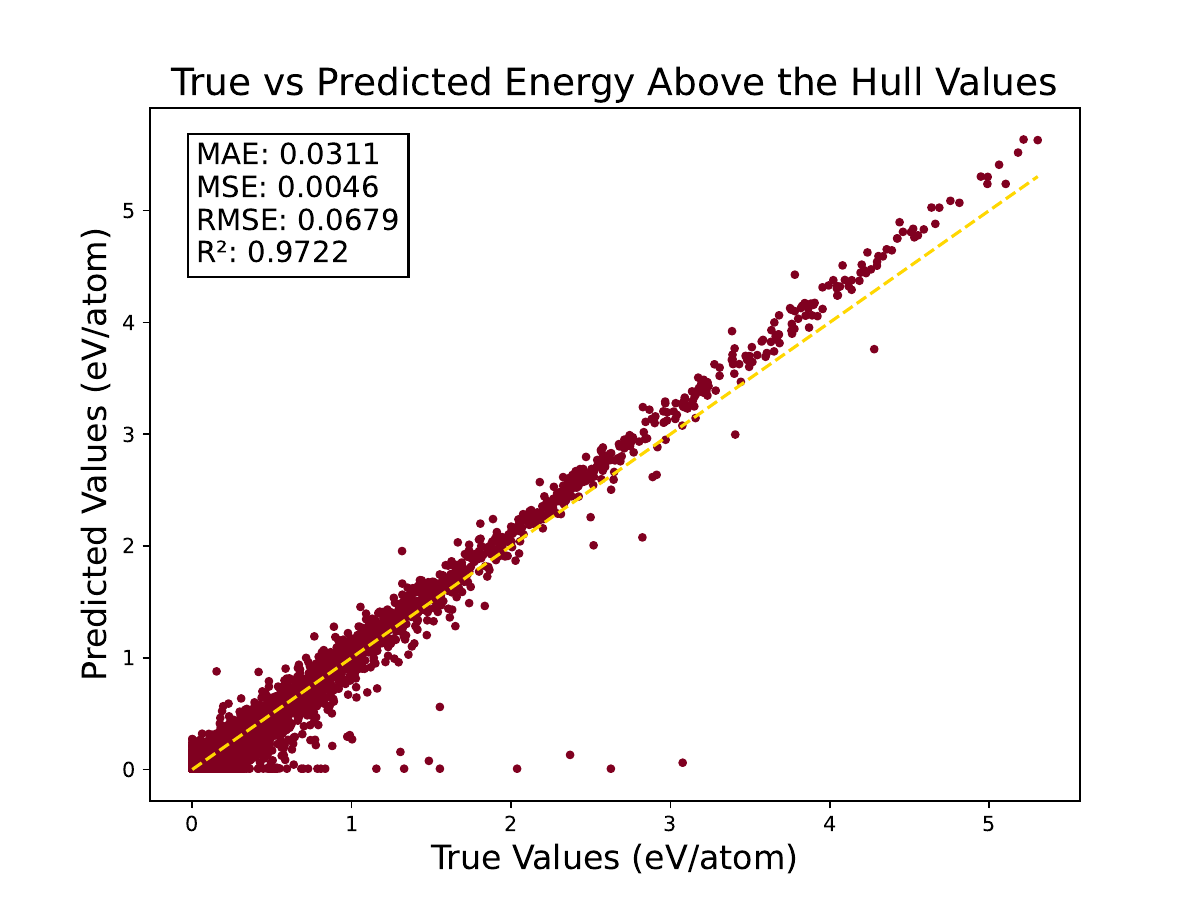}
	\caption{True values vs.\ predicted values for energy above the hull. This plot shows the true values vs.\ predicted values for energy above the hull for materials with corresponding space group classifications. The following error metrics were calculated: MAE = 0.0311 eV/atom, MSE = 0.0046 eV/atom, RMSE = 0.0679 eV/atom, $R^2$ = 0.9722.}
	\label{fig:ehull}
\end{figure}

After developing a model that predicts formation energy, we used the same architecture to predict energy above the hull, with formation energy included as an additional input feature. We use only the space group classifications in the input feature. Generally, compounds with zero energy above the hull are considered stable. However, in this study, we will also consider compounds with energy above the hull that is almost zero which can be considered as metastable. Early stopping terminated the training after epoch 152. Figure \ref{fig:ehull} shows the scatter plot of true values vs.\ predicted versus for energy above the hull during the testing phase.

\subsection{Stability Prediction of Manganese-Nickel-Oxygen Chemical System}
We now have two models: one that predicts formation energy and another that predicts energy above the hull. We systematically generated potential ternary compounds of the Manganese-Nickel-Oxygen chemical system, combining Manganese, Nickel, and Oxygen using the itertools library \cite{van1995python}. By considering possible subscripts ranging from 1 to 8 for each element, we permuted and assigned these numbers to every conceivable combination to construct compounds. Subsequently, we assigned each generated compound to all possible space group symmetries, resulting in a total of 77,280 combinations. We removed compounds assigned to space groups 168 and 207, as these space groups do not exist in the Materials Project data, reducing the total to 76,608 entries. However, the unique compounds numbered only 336. Using Pymatgen, we converted the compounds into their chemical compositions and then featurized them into elemental fractions with Matminer \cite{ward2018matminer}.

To ensure both novelty and diversity in our dataset, we meticulously filtered out compounds already present in our training and testing sets. As a result, we successfully removed 33 compounds, of which only 9 were unique. Some of the removed compounds were duplicates but had different space groups.

Finally, leveraging our developed model, we determined the most stable space group symmetry for each of the newly generated compounds from this meticulously curated dataset. Table \ref{tab:material_properties} shows the material properties and predicted formation energy and energy above hull values in eV/atom for various compounds in the Manganese-Nickel-Oxygen chemical system, including their space group symmetries, predicted formation energies, and energy above hull values. 

\begin{table}[h!]
	\centering
	\renewcommand{\arraystretch}{1.2} \extrarowheight0.5ex
	\setlength{\tabcolsep}{1.1ex}
	\begin{tabular}{cccc}
		\hline \hline
		\multirow{2}{*}{material} & \multirow{2}{*}{space group} & predicted formation energy& predicted E$_{\text{hull}}$\\
		&& (eV/atom) & (eV/atom) \\
		\hline
		Mn$_2$Ni$_3$O	 & 103	 & -2.223632 	 & 0.007018 \\
		Mn$_2$Ni$_3$O$_4$ & 103   & -2.419526 	 & 0.007018 \\
		Mn$_2$Ni$_3$O$_5$ & 214   & -2.419526 	 & 0.007018 \\
		Mn$_2$Ni$_3$O$_6$ & 196   & -2.468194 	 & 0.007018 \\
		Mn$_2$Ni$_3$O$_7$ & 196   & -2.478734 	 & 0.007018 \\
		\hline \hline
	\end{tabular}
	\caption{Generated chemical compound materials and their predicted formation energy and energy above hull.}
	\label{tab:material_properties}
\end{table}

\section{Conclusions}
\label{sec:conclusions}

We have successfully designed a deep neural network model that predicts formation energy using chemical compositions along with integrated symmetry classifications input features. Incorporating symmetry classifications led to improved results, with space group information contributing the most to the accuracy enhancement, followed by point group information, and then crystal system information. Additionally, we developed a model to forecast the energy above the hull, a key indicator of material stability. By incorporating the symmetry classifications, our approach provides insights into the potential crystal symmetries that specific chemical compounds might exhibit. This demonstrates the critical role of crystallographic data in enhancing the predictive capabilities of deep learning models for material properties. 

\section*{Acknowledgments}
We appreciate stimulating discussions with H.~Aringa, V.~Convicto, and M.~Dengal.

\section*{Author contributions}

\textbf{V.\ Torlao:} Conceptualization, Data curation, Formal analysis, Investigation, Methodology, Resources, Software, Validation, Visualization, Writing - original draft, Writing - review and editing.
\textbf{E. A. Fajardo:} Conceptualization, Methodology, Project administration, Supervision, Writing - review and editing.

\section*{Declaration of interests}

The authors declare no financial interests and personal relationships that could influence or bias the work reported in this paper.

\section*{Funding sources}

This work is supported by the Science and Technology Regional Alliance of Universities for National Development (STRAND) scholarship granted to V.\ Torlao by the Department of Science and Technology-Science Education Institute (DOST-SEI).

\bibliographystyle{elsarticle-num} 

\begin{thebibliography}{10}
\expandafter\ifx\csname url\endcsname\relax
  \def\url#1{\texttt{#1}}\fi
\expandafter\ifx\csname urlprefix\endcsname\relax\def\urlprefix{URL }\fi
\expandafter\ifx\csname href\endcsname\relax
  \def\href#1#2{#2} \def\path#1{#1}\fi

\bibitem{kirklin2015open}
S.~Kirklin, J.~E. Saal, B.~Meredig, A.~Thompson, J.~W. Doak, M.~Aykol, S.~R{\"u}hl, C.~Wolverton, The open quantum materials database (oqmd): assessing the accuracy of dft formation energies, npj Computational Materials 1~(1) (2015) 1--15.

\bibitem{sun2016thermodynamic}
W.~Sun, S.~T. Dacek, S.~P. Ong, G.~Hautier, A.~Jain, W.~D. Richards, A.~C. Gamst, K.~A. Persson, G.~Ceder, The thermodynamic scale of inorganic crystalline metastability, Science advances 2~(11) (2016) e1600225.

\bibitem{jain2013commentary}
A.~Jain, S.~P. Ong, G.~Hautier, W.~Chen, W.~D. Richards, S.~Dacek, S.~Cholia, D.~Gunter, D.~Skinner, G.~Ceder, et~al., Commentary: The materials project: A materials genome approach to accelerating materials innovation, APL materials 1~(1) (2013).

\bibitem{saal2013materials}
J.~E. Saal, S.~Kirklin, M.~Aykol, B.~Meredig, C.~Wolverton, Materials design and discovery with high-throughput density functional theory: the open quantum materials database (oqmd), Jom 65 (2013) 1501--1509.

\bibitem{ward2018matminer}
L.~Ward, A.~Dunn, A.~Faghaninia, N.~E.~R. Zimmermann, S.~Bajaj, Q.~Wang, J.~H. Montoya, J.~Chen, K.~Bystrom, M.~Dylla, et~al., Matminer: An open source toolkit for materials data mining, Computational Materials Science 152 (2018) 60--69.

\bibitem{ye2018pauling}
W.~Ye, C.~Chen, Z.~Wang, I.-H. Chu, S.~P. Ong, Deep neural networks for
accurate predictions of crystal stability, Nature communications 9~(1) (2018) 3800.

\bibitem{jha2018elemnet}
D.~Jha, L.~Ward, A.~Paul, W.-k. Liao, A.~Choudhary, C.~Wolverton, A.~Agrawal, Elemnet: Deep learning the chemistry of materials from only elemental composition, Scientific reports 8~(1) (2018) 17593.

\bibitem{goodall2020roost}
R.~E. Goodall, A.~A. Lee, Predicting materials properties without crystal structure: Deep representation learning from stoichiometry, Nature communications 11~(1) (2020) 6280.

\bibitem{jain2018atomic}
A.~Jain, T.~Bligaard, Atomic-position independent descriptor for machine learning of material properties, Physical Review B 98~(21) (2018) 214112.

\bibitem{ward2017including}
L.~Ward, R.~Liu, A.~Krishna, V.~I. Hegde, A.~Agrawal, A.~Choudhary, C.~Wolverton, Including crystal structure attributes in machine learning models of formation energies via voronoi tessellations, Physical Review B 96~(2) (2017) 024104.

\bibitem{kim2018machine}
K.~Kim, L.~Ward, J.~He, A.~Krishna, A.~Agrawal, C.~Wolverton, Machine-learning-accelerated high-throughput materials screening: Discovery of novel quaternary heusler compounds, Physical Review Materials 2~(12) (2018) 123801.

\bibitem{aggarwal2018neural}
C.~C. Aggarwal, et~al., Neural networks and deep learning, Springer 10~(978) (2018) 3.

\bibitem{glorot2011deep}
X.~Glorot, A.~Bordes, Y.~Bengio, Deep sparse rectifier neural networks, in: Proceedings of the fourteenth international conference on artificial intelligence and statistics, JMLR Workshop and Conference Proceedings, 2011, pp. 315--323.

\bibitem{kingma2015adam}
D.~Kingma, J.~Ba, Adam: A method for stochastic optimization, in: International Conference on Learning Representations (ICLR), San Diega, CA, USA, 2015.

\bibitem{duchi2011adaptive}
J.~Duchi, E.~Hazan, Y.~Singer, Adaptive subgradient methods for online learning and stochastic optimization., Journal of machine learning research 12~(7) (2011).

\bibitem{tieleman2012root}
T.~Tieleman, G.~Hinton, Root mean square propagation. divide the gradient by a running average of its recent magnitude. coursera: Neural networks for machine learning (2012).

\bibitem{van1995python}
G.~Van~Rossum, The Python Library Reference, release 3.8.2, Python Software Foundation, 2020.

  
\end{thebibliography}

\end{document}